\documentclass[12pt]{article}
\usepackage[T1]{fontenc}
\usepackage{textcomp}
\usepackage{mathptmx}

\usepackage{graphicx}
\usepackage{amsmath,amsthm,amssymb}
\newtheorem{remark}{Remark}
\newcommand{\san}[1]{\mathsf{#1}}
\newtheorem{theorem}{Theorem}
\newtheorem{lemma}{Lemma}
\newtheorem{corollary}{Corollary}

\newcommand{\ket}[1]{|#1\rangle}
\newcommand{\bra}[1]{\langle #1 |}

\title{Secure Key Rate of the BB84 Protocol using Finite Sample Bits\thanks{%
To be published in J.\ Phys.\ A.
Five page abstract of this paper appeared in Proc.\ 2010 IEEE International Symposium on Information Theory, June 13--18, 2010,  Austin, Texas, USA.}}
\author{Yousuke Sano, Ryutaroh Matsumoto\footnote{ryutaroh@rmatsumoto.org}, Tomohiko Uyematsu\footnote{uyematsu@ieee.org}\\
\normalsize{Department of Communications and Integrated Systems}\\
\normalsize{Tokyo Institute of Technology}\\
\normalsize{Oookayama, Meguro-ku Tokyo, 152--8552, Japan}}
\date{October 2010}
\begin{document}
\maketitle

\begin{abstract}
We improve the non-asymptotic key rate shown by Scarani and Renner by proposing several methods to construct tighter conservative confidence intervals of the phase error rate than one shown by them. In addition, we show that the accurate channel estimation method non-asymptotically increases the key rate over the amplitude damping channel as well as the asymptotic case in the BB84 protocol.
\end{abstract}

\section{Introduction}
Quantum key distribution (QKD) has attracted great attention
as a feasible application of quantum information science with the current device technology \cite{donna}. The goal of a QKD protocol is to share a random bit sequence not known by the eavesdropper Eve, between the legitimate sender Alice and the receiver Bob. The fundamental feature of QKD protocols is that the maximum amount of information gained by Eve can be determined from the channel estimate between Alice and Bob. Such a task cannot be conducted in classical key distribution schemes. If the estimated amount is lower than a threshold, then Alice and Bob determine the length of a secret key
from the estimated amount of Eve's information, and can share the secret key by performing the information reconciliation (error correction) and the privacy amplification. Since the key rate, which is the length of securely sharable key per channel use, is one of the most important criteria for the efficiency of QKD protocols, the estimation of the channel is of primary importance.

Conventionally in the  Bennett-Brassard 1984 (BB84) protocol \cite{bb84}, we only use the statistics of matched measurement outcomes which are transmitted and received by the same basis, to estimate the quantum channel; mismatched measurement outcomes, which are transmitted and received by different basis, are discarded in the conventionally used channel estimation methods. By contrast, Watanabe \textit{et al.}\ \cite{watanabe:08} showed that by using the statistics of mismatched measurement outcomes in addition to that of matched measurement outcomes, we can estimate a quantum channel more accurately, thereby a higher key rate can be achieved than the conventional one. However their analysis was only asymptotic, i.e., they assumed that the number of sample bits for channel estimation is infinite. Hence, for practical use, it is necessary to perform non-asymptotic analysis.

For non-asymptotic analysis of the QKD protocol, Scarani \textit{et al.}\ formulated a lower bound on secure key rate \cite{PRL,scarani:08}. Other researches of non-asymptotic analysis was surveyed by Cai \textit{et al.}\ \cite{cai:08}. Since the formula by Scarani \textit{et al.}\ has enough generality, in theory it enables us to calculate not only non-asymptotic key rate based on the conventional channel estimation but also the one based on the accurate channel estimation \cite{watanabe:08} for the BB84 protocol. 
%Nevertheless, an analytical computation of the former may be difficult. 

On the other hand, in Cai \textit{et al.}\ \cite[p.4]{cai:08}, it was suggested that a lower bound on secure key rate shown by Scarani \textit{et al.}\ might be able to be improved. In the channel estimation step shown by Scarani \textit{et al.}\, the channel parameter is guessed by \textit{interval estimation}. However, the method of constructing confidence region of the interval estimation is not unique. Even when we use the confidence region which is different from the one shown by Scarani \textit{et al.}\, if it satisfies the condition of \textit{conservativeness}, the security of the final key is guaranteed. Specifically, even if \textit{one-sided interval estimation} is used, the security is still kept.

In this paper, we show two things: First, we show several methods of reconstructing the confidence region, and the fact that they increase the non-asymptotic secure key rate in the BB84 protocol. Second, we show the utility of accurate channel method on the BB84 protocol using finite sample bits. To do this, we compare the non-asymptotic key rate based on the accurate channel estimate to the  conventional one by numerical computation over the amplitude damping channel and the depolarizing channel. 

We stress that the assumption used in this paper is exactly the same as
\cite{PRL}. In particular, we assume no prior knowledge of channel
nor channel model
with the accurate channel esitimation, as well as its asymptotic case
\cite{watanabe:08}.
In the numerical comparison in Section \ref{sec:numcompt},
we shall use the depolarizing channel and the amplitude damping channel
to generate the mesurement outcomes, but the proposed protocols do not
assume the knowledge of the underlying channels, and estimate the channel
among all the possible channels.

The rest of this paper is organized as follows:
We first review previously known results in Section \ref{section-pre}. Second, we show several methods to improve the key rate and the results of the improvements in Section \ref{section-imp}. Last, we state the conclusion in Section \ref{section-conclusion}.

\section{Preliminaries}
\label{section-pre}
\subsection{BB84 protocol}
Typical one of the QKD protocols is the BB84 protocol invented by Bennett and Brassard \cite{bb84}. The goal of the BB84 protocol is to share a random bit sequence not known by the eavesdropper Eve, between the legitimate sender Alice and the receiver Bob. In the following, we briefly describe the flow of the protocol and the accurate channel estimation shown by Watanabe \textit{et al.}\ \cite{watanabe:08} on the BB84 protocol.
\subsubsection{Overview of BB84 protocol}
BB84 protocol consists of the following four steps:
\begin{enumerate}
 \item \textit{Distribution of quantum information: } Alice sends $N$ quantum objects, for example photon polarizations, to Bob over a quantum channel.
 \item \textit{Parameter (or channel) estimation: }Alice and Bob disclose a part of the transmission/received information to each other to estimate the quantum channel between Alice and Bob.\label{step2}
 \item \textit{Information reconciliation: }For the bit string not disclosed in step \ref{step2}, Alice sends the syndrome to Bob, and Bob corrects an error by using the syndrom. \label{step3}
 \item \textit{Privacy amplification: }Alice and Bob compress corrected bit strings in step \ref{step3} according to the same hash function so that compressed bit string is statistically independent of information obtained by Eve. Consequently, compressed bit string is the final key.
\end{enumerate}
The security of the final key obtained by BB84 protocol can be proven only by the axiom of quantum mechanics \cite{PRL,scarani:08}.

\subsubsection{Accurate channel estimation}
\label{section-accurate}
In this section, we explain the distribution of quantum informationin and convenational parameter estimation more concretely. Moreover, we explain the accurate channel estimation shown by Watanabe et al. [3] on the BB84 protocol.

Alice first randomly sends bit $0$ or $1$ to Bob by modulating it into a transmission basis that is randomly chosen from
the $\san{z}$-basis $\{ \ket{0_\san{z}}, \ket{1_\san{z}} \}$,
the $\san{x}$-basis $\{ \ket{0_\san{x}}, \ket{1_\san{x}} \}$, where $\ket{0_a},\ket{1_a}$ are eigenstates of the Pauli matrix $\sigma_a$ for $a\in\{\san{z},\san{x}\}$, respectively.
Then Bob randomly chooses one of measurement observables
$\sigma_\san{z}$, $\sigma_\san{x}$, and converts
a measurement result $+1$ or $-1$ into
a bit $0$ or $1$ respectively.
After $N$ transmissions, Alice
and Bob publicly announce their transmission bases and
measurement observables. 
They also announce $m(<N)$ bits of their bit sequence for
estimating channel $\mathcal{E}_B$ from Alice to Bob.
Conventionally, Alice and Bob discard mismatched measurement outcomes,
which are transmitted and received by different bases \cite{shor}. In contrast, Watanabe \textit{et al.}\ \cite{watanabe:08} show that by using the statistics of mismatched measurement outcomes in addition to that of matched measurement outcomes, we can estimate a quantum channel more accurately, thereby the key rate is at least higher than the conventional one. In particular, the key rate is generally improved over the conventional one over any channel, and only if the quantum channel is the Pauli channel, those two key rates are equal \cite{watanabe:09}.

%******************************************************************************
\subsection{Method of type}
\label{type}
In this section, we review the method of type \cite[Chapter 11]{cover:01}
that are used in this paper. Let $\cal{X}$ be a finite set. For a sequence $x^m = (x_1, \ldots, x_m) \in {\cal X}^m$,
the type of $x^m$ is the 
empirical probability distribution $P_{x^m}$ defined by
\begin{eqnarray*}
P_{x^m}(a) := \frac{ | \{ i \mid x_i = a \} | }{m}~~~~~~\mbox{for }
a \in {\cal X}.
\end{eqnarray*}
Then, the following theorems hold. 
\begin{theorem}\label{sebsection-theorem1}
\textnormal{\cite[Theorem 11.2.1]{cover:01}\;
Let $P_X$ be a probability distribution on $\mathcal{X}$ and $P_{x^m}$ be
the type of the sequence $x^m$ drawn according to the $m$-fold product distribution $P_X^m$. Then, for any
$\delta$,
\begin{equation*}
\textnormal{Pr}\bigl[ D(P_{x^m}|| P) > \delta \bigl] \leq 2^{-m(\delta-|\mathcal{X}|\frac{\log_2(m+1)}{m})}
\end{equation*}
where $D(\cdot)$ is the relative entropy. Note that the base of a logarithm and a (conventional) entropy are $2$ throughout this paper.}
\end{theorem}

\begin{lemma} \label{lemma}
\textnormal{\cite[Theorem 11.6.1]{cover:01}\;
Let $P$ and $Q$ be probability distributions. Then
\begin{equation*}
||P - Q||_1 \leq \sqrt{2(\ln2)D(P || Q)}
\end{equation*}
where $||\cdot||_1$ is the variational distance defined by $||P_1-P_2||_1:=\sum_{x\in\mathcal{X}} |P_1(x)-P_2(x)|$, where $P_1,P_2$ are probability mass functions on $\mathcal{X}$.}
\end{lemma}

\begin{corollary}\label{coro10}
\textnormal{
Let $P_{x^m}$ be the type of $X$. For any $\delta>0$,
\begin{equation*}
\mathrm{Pr}\bigl[ ||P_{x^m} - P||_1 > \delta \bigl] \leq 2^{-m(\frac{\delta^2}{2\ln{2}}-|\mathcal{X}|\frac{\log_2(m+1)}{m})}.
\end{equation*}
}
\end{corollary}

%\vspace{-1.5mm}
\subsection{Non-asymptotic key rate analysis}
%\vspace{-1.5mm}
In this section, we rephrase non-asymptotic key rate analysis shown by Scarani \textit{et al.}\ \cite{scarani:08} on the BB84 protocol in terms of interval estimation. This paraphrase is necessary to clarify the relation between the method shown by Scarani \textit{et al.}\ and our proposed one. Note that interval estimation of a quantum channel for QKD protocols is also discussed in \cite{leverrier:09,leverrier10}.

%In the Stokes parameterization, the qubit channel $\mathcal{E}_B$
%can be described by the affine map parameterized
%by $12$ real parameters \cite{fujiwara:98,fujiwara:99}:
%\begin{eqnarray}
%\left[ \begin{array}{c}
%\theta_{\san{z}} \\ \theta_{\san{x}} \\ \theta_{\san{y}}
%\end{array} \right] 
%\mapsto
%\left[ \begin{array}{ccc}
%R_{\san{zz}} & R_{\san{zx}} & R_{\san{zy}} \\
%R_{\san{xz}} & R_{\san{xx}} & R_{\san{xy}} \\
%R_{\san{yz}} & R_{\san{yx}} & R_{\san{yy}}
%\end{array} \right]
%\left[ \begin{array}{c}
%\theta_{\san{z}} \\ \theta_{\san{x}} \\ \theta_{\san{y}}
%\end{array} \right]
%+ \left[ \begin{array}{c}
%t_{\san{z}} \\ t_{\san{x}} \\ t_{\san{y}} 
%\end{array} \right],
%\label{eq-affine-map}
%\end{eqnarray}
%where $(\theta_\san{z}, \theta_\san{x}, \theta_\san{y})$
%describes a vector in the Bloch sphere \cite{nielsen-chuang:00}.

%******************************************************************************
%******************************************************************************

\subsubsection{Interval estimation}
Here we briefly review some basic concepts of the interval estimation. See textbooks of statistics for more details (e.g. \cite{casella-berger:01}).

The goal of the interval estimation is to estimate the unknown statistical parameter $\theta$ by observed samples. First, we define the \textit{confidence region}. Let a sample sequence $X=X_1,\cdots,X_n\sim P_{\theta}$ be \textit{i.i.d.}, and $\Theta$ be the parameter space. For any $\alpha$ between $0$ and $1$, if a set $C(X) \subset\Theta$ satisfies 
%\vspace{-2mm} 
\begin{equation}
\forall\theta\in\Theta, P_{\theta}[\theta\in C(X)] \gtrsim 1-\alpha, \label{intereval-region}
\end{equation}
then $C(X)$ is called a \textit{confidence region}. Specially, if $\theta$ is real-valued, then $C(X)$ is usually an interval of real numbers, sometimes called the \textit{confidence interval}. In addition, the real-number $1-\alpha$ is called the \textit{confidence level} or \textit{confidence coefficient}. If the inequality in Eq.~(\ref{intereval-region}) is always satisfied, i.e.
\begin{equation*}
\forall\theta\in\Theta, P_{\theta}[\theta\in C(X)] \geq 1-\alpha,
\end{equation*}
then such $C(X)$ is called a \textit{conservative} confidence region. 
%Interval estimation may be considered as concstructing the tight confidence region to required confidence level.

Second, we describe the \textit{one-sided interval estimation}. Suppose that $\theta$ is a real number. One-sided interval estimation is defined as constructing the upper bound on $\theta$ satisfying
\begin{equation}
\forall\theta\in\Theta, P_{\theta}[\theta \leq C(X)] \gtrsim 1-\alpha. \label{one-sided}
\end{equation}
The interval $(-\infty,C(X)]$ is called a \textit{one-sided confidence interval} with confidence level $1-\alpha$. Of course, if the inequality in Eq.~(\ref{one-sided}) is always satisfied, the interval $(-\infty,C(X)]$ is conservative.

%******************************************************************************
\subsubsection{Channel estimation using finite sample bits}
One of the practical issues of QKD protocol is that sample bits used for channel estimation is limited to a finite number. Scarani \textit{et al.}\ showed a method for interval estimation of the quantum channel \cite{PRL,scarani:08}. Hereafter, the basis $\{\ket{0},\ket{1}\}$ is  the $\san{z}$-basis unless otherwise stated.

The channel $\mathcal{E}_B$, which denotes a qubit channel from Alice to Bob, can be also described by the Choi operator \cite{choi} $\rho_{AB} := (id\otimes\mathcal{E}_B)(\ket{\psi}\bra{\psi})$ for the Bell state $\ket{\psi}=\frac{1}{\sqrt{2}}(\ket{0}\ket{0}+\ket{1}\ket{1})$. For any $\epsilon_{PE} (0 \leq \epsilon_{PE} \leq 1)$, let \cite{cai:08}  
\begin{eqnarray}
\xi &:=& \sqrt{\frac{2\ln{(1/\epsilon_{PE})}+2d\ln{(m+1)}}{m}}, \nonumber \\
\Gamma_{\xi} &:=& \bigl\{ \rho_{AB} : ||\lambda_m - \lambda_{\infty}(\rho_{AB})||_1 \leq \xi \bigl\}, \label{variational}
\end{eqnarray}
where $\lambda_m$ are obtained by measurements of $m$ samples of $\rho_{AB}$ according to a POVM measurement with $d$ outcomes, and $\lambda_{\infty}(\rho_{AB})$ denotes the perfect statistics in the limit of infinitely many measurements.
Then $\Gamma_{\xi}$ can be interpreted as the conservative confidence region with confidence level $1-\epsilon_{PE}$ for the qubit channel $\rho_{AB}$. Indeed, for any $\rho_{AB}$, we can see 
\begin{eqnarray}
\textnormal{Pr}[||\lambda_m - \lambda_{\infty }(\rho_{AB})||_1 \leq \xi] &\geq& 1-2^{-m(\frac{\xi^2}{2\ln{2}}-d\frac{\log_2{(m+1)}}{m})}\label{eq100}\\
&=&1-\epsilon_{PE}\nonumber
\end{eqnarray}
by Corollary \ref{coro10} in Section \ref{type}.
Note that the definition of variational distance used in this paper
is the same as \cite{cover:01} and twice as large as the one used in \cite{cai:08} and that
the right hand side of Eq.\ (\ref{eq100}) is twice as large as
\cite[Eq.\ (3)]{cai:08}, where \cite[Eq.\ (3)]{cai:08}
is corrected in its erratum.
By the same argument as \cite{cai:08},
we see that for $d=2$ we can use
\begin{equation}
p_\mathrm{observed}+ \sqrt{\frac{\ln{(1/\epsilon_{PE})}+2\ln{(m+1)}}{2m}}\label{eq101}
\end{equation}
as the worst-case estimate of the so-called phase error rate,
where $p_\mathrm{observed}$ is the actually observed phase error rate.
We shall use Eq.\ (\ref{eq101}) in the numerical comparison in Section
\ref{sec:numcompt}.

%******************************************************************************
\subsubsection{Lower bound on the secure key rate of the BB84 protocol}
First, we define \textit{$\epsilon$-security} \cite{ben:04,renner:04}. For any $\epsilon \geq 0$, a final key $K$ is said to be \textit{$\epsilon$-secure} with respect to an adversary Eve if the joint state $\rho_{KE}$ satisfies
\begin{equation*}
||\rho_{KE}-\tau_{K}\otimes\rho_{E}|| \leq \epsilon,
\end{equation*}
where $\tau_K$ is the completely mixed state on a key space $\mathcal{S}_K$, and $||\cdot||$ is the trace distance.
The parameter
$\epsilon$ can be interpreted as the maximum failure probability in which an adversary might have gained some information on $K$.

Next, we describe the lower bound on the $\epsilon$-secure key rate of the BB84 protocol using finite samples shown by Scarani \textit{et al.}\ \cite{scarani:08}. If the length $l$ of the final key is 
\begin{equation}
 l = N \bigl[ \min_{\rho_{AB}\in\Gamma_{\xi}}S_{\rho_{AB}}(X|E)-\delta(\bar{\epsilon}) \bigl] -\textnormal{leak}_{\epsilon_{EC}}-2\log_2{\frac{1}{\epsilon_{PA}}}, \label{lower-bound}
  \end{equation}
then the final key is \textit{$\epsilon$-secure}, where $S_{\rho_{AB}}(X|E)$ is the conditional von Neumann entropy for the state $\rho_{AB}$,
% whose base of the logarithm is 2, 
and $\Gamma_{\xi}$ is the confidence region for $\rho_{AB}$ with the confidence level $1-\epsilon_{PE}$, and $\epsilon\geq\epsilon_{PE}$. See \cite{PRL,scarani:08} for more detail of Eq.~(\ref{lower-bound}). This formula enables us to calculate the non-asymptotic key rate based on the accurate channel estimate and the conventional one for the BB84 protocol respectively.
%\vspace{-2mm} 
\begin{remark}
\textnormal{Eve's ambiguity for Alice's bit $S_{\rho_{AB}}(X|E)$
can be calculated from the Choi operator $\rho_{AB}$ as follows. 
Let the density operator $\rho_{XB}$ be derived by measurement on Alice's system, i.e., $\rho_{XB}:=\sum_{x\in\mathbb{F}_2}(\ket{x}\bra{x}\otimes I)\rho_{AB}(\ket{x}\bra{x}\otimes I)$. The conditional von Neumann entropy $S_{\rho_{AB}}(X|E)$ is defined by $S_{\rho_{AB}}(X|E):=S(\rho_{XE})-S(\rho_E)$, where $S(\cdot)$ is the von Neumann entropy and $\rho_{E}$ is the partial trace of $\psi_{ABE}$, which is the purification of $\rho_{AB}$, over the joint system of Alice and Bob. Noting that $S_{\rho_{AB}}(X|E)=S_{\rho_{AB}}(X|B)$ and $S(\rho_{E})=S(\rho_{AB})$, $S{\rho_{AB}}(X|E)$ can be calculated by $S_{\rho_{AB}}(X|E)=S(\rho_{XB})-S(\rho_{AB})$.
%Since system $X$ is classical, we can rewrite 
%$S(\rho_{XE}) = H(X) + \sum_{x \in \mathbb{F}_2} \frac{1}{2} 
%S(\mathcal{E}_E(\ket{x}\bra{x}))$, where $H(X)$ is the shannon entropy with res%pect to $X$, and $\mathcal{E}_E$ denotes the channel to environment of channel $\mathcal{E}_B$.
%Noting that  $S(\mathcal{E}_E(\ket{x}\bra{x})) = S(\mathcal{E}_B(\ket{x}\bra{x}))$
%and $S(\rho_E) = S(( id \otimes \mathcal{E}_B)(\psi))$ for
%the Bell state $\ket{\psi}$, $S_{\rho_{AB}}(X|E)$ can be calculated from the channel $\mathcal{E}_B$
}
\end{remark}
%\vspace{-1mm} 

%******************************************************************************
%******************************************************************************
%\vspace{-2mm}
\section{Improvement of key rate}
%\vspace{-2mm}
\label{section-imp}
In this section, we present several methods of improving the lower bound on the secure key rate by replacing the confidence region as shown in Eq.~(\ref{variational}). In general, the smaller the confidence region $\Gamma_{\xi}$ is, the bigger Eve's worst-case ambiguity $\min_{\Gamma_{\xi}}S_{\rho_{AB}}(X|E)$ can grow. Even when we use the confidence region which is different from $\Gamma_{\xi}$, if it is conservative, the security of the final key is guaranteed. Hence, the lower bound in Eq.~(\ref{lower-bound}) can be improved by reconstructing the confidence region with confidence level $1-\epsilon_{PE}$ tighter than $\Gamma_{\xi}$ because the influence from the different channel estimation method appears only in Eve's worst-case ambiguity in Eq.~ (\ref{lower-bound}). In addition, we clarify the utility of the accurate channel estimation method in the BB84 protocol using finite sample bits by numerically computing Eve's worst-case ambiguities over the amplitude damping channel and the depolarizing channel. 

We first present several methods for composing such confidence region in Section \ref{re-confi}. Then we show how to compute Eve's worst-case ambiguity with the accurate channel estimation in Section \ref{section-computing}. Last we compare Eve's worst-case ambiguities by the proposed methods and the accurate channel estimation in Section \ref{section-comparison}.  Hereafter, we distinguish the conventional channel estimation reviewed Section \ref{section-accurate} and the conventional confidence region shown by Scarani \textit{et al.}\ \cite{cai:08} to avoid confusion. We call the former the conventional channel estimation, and the latter the conventional confidence region or merely $\Gamma_{\xi}$.

%\vspace{-2.5mm}
\subsection{Reconstruction of confidence region}
\label{re-confi}
%******************************************************************************
%\vspace{-1.5mm} 
\subsubsection{Relative entropy}
\label{subsection-relative-entropy}
Here, we reconstruct the confidence interval with confidence level $1-\epsilon_{PE}$ using the relative entropy. Let
%\vspace{-2.5mm} 
\begin{eqnarray}
\xi' &:=& \frac{\log_2{(1/\epsilon_{PE})+d\log_2{(m+1)}}}{m}, \nonumber \\
\Gamma_{\xi'} &:=& \bigl\{ \rho_{AB} : D(\lambda_m||\lambda_{\infty}(\rho_{AB})) \leq \xi' \bigl\}, \label{relative}
\end{eqnarray}
where $D(\cdot)$ is the relative entropy \cite{cover:01}.
Then in the following, we prove that the set $\Gamma_{\xi'}$ is the conservative confidence region for $\rho_{AB}$ with confidence level $1-\epsilon_{PE}$, and $\Gamma_{\xi'}\subset\Gamma_{\xi}$. 

%Those can be easily proved by Theorems 11.2.1 and 11.6.1 of \cite{cover:01}. Hence, by replacing $\Gamma_{\xi}$ of Eq.~(\ref{lower-bound}) with $\Gamma_{\xi'}$, we can surely gain a higher key rate than the conventional one.
\begin{proof}
From Theorem \ref{sebsection-theorem1}, obviously
\begin{eqnarray*}
\textnormal{Pr} \bigl[ D(\lambda_m || \lambda_{\infty}(\rho_{AB}) ) \leq \xi' \bigl] &\geq& 1-2^{-m(\xi'-d\frac{\log_2(m+1)}{m})}\\
&=&1-\epsilon_{PE}
\end{eqnarray*}
Thus, $\Gamma_{\xi'}$ is the conservative confidence region for $\rho_{AB}$ with confidence level $1-\epsilon_{PE}$. In addition, let 
\begin{equation*}
\eta=D(\lambda_m || \lambda_{\infty}(\rho_{AB})) - \frac{||\lambda_m - \lambda_{\infty}(\rho_{AB})||^2_1}{2\ln{2}},
\end{equation*}
then 
\begin{eqnarray*}
& & ||\lambda_m - \lambda_{\infty}(\rho_{AB})||_1 \leq \xi\\
&\Leftrightarrow& D(\lambda_m || \lambda_{\infty}(\rho_{AB})) \leq \xi'+\eta
\end{eqnarray*}
Thus, $\Gamma_{\xi}$ can be rewritten as follows; 
\begin{equation*}
\Gamma_{\xi} = \bigl\{ \rho_{AB} : D(\lambda_m||\lambda_{\infty}(\rho_{AB})) \leq \xi' + \eta  \bigl\}.
\end{equation*}
From Lemma \ref{lemma}, $\eta\geq 0$. Therefore, $\Gamma_{\xi'} \subseteq \Gamma_{\xi}$.
\end{proof}
Hence, by replacing $\Gamma_{\xi}$
of Eq.~ (\ref{lower-bound}) with $\Gamma_{\xi'}$ , we can surely gain a higher key rate than
the conventional one.

%******************************************************************************
%Thus, we propose here sereral methos for the one-sided interval estimation \cite{casella-berger:01} of phase error rate, i.e. for constructing 
%, i.e. into conservative upper bound $C(X)$ like Eq.~(\ref{one-sided})

\subsubsection{Binomial one-sided confidence bounds}
\label{section-bino}
Here we describe a general method for converting an upper bound on the tail probability of the binomial distribution $B(m,p)$ into the conservative one-sided confidence interval for $p$ with confidence level $1-\epsilon_{PE}$, where $m$ is the number of Bernoulli trials and $p$ is the probability of success on each trial. In the conventional channel estimation, Eve's worst-case ambiguity can be calculated by the estimated phase error rate \cite{renner:05}. We can use the one-sided interval estimation \cite{casella-berger:01} to guess the phase error rate. The one-sided interval estimation for phase error rate is equivalent to that for $p$ of binomial distribution $B(m,p)$, which can be performed by converting an upper bound on the tail probability of $B(m,p)$. Thus we describe such a method. In addition, we enumerate concretely some upper bounds for $B(m,p)$, and show that one-sided confidence intervals gained by those bounds can increase Eve's worst-case ambiguity compared with Eqs.~(\ref{variational}) and (\ref{relative}).
%Eve's worst-case ambiguity can be calculated by estimated phase error rate in the conventional BB84 protocol \cite{renner:05}. We can use the one-sided interval estimation \cite{casella-berger:01} to guess the phase error rate. Furthermore, the estimation of the phase errro rate $p$ is equivalent to parameter estimation of the binomial distribution $B(m,p)$, where $m$ is the number of trials. An upper bound on tail probability of the binomial distribution can generally convert to the one-sided confidence interval. Therefore, we describe a general converting method and some upper bounds, which can be converted to the one-sided confidence interval for $p$ with confidence level $1-\epsilon_{PE}$. 

Hereafter, $X$ be a random variable according to $P_X=B(m,p)$, and $\bar{X}=X/m$.
\begin{enumerate}
 \item \textit{Preliminary}
 \label{section-preliminaly}\;:\;First of all, we describe the general converting method. Our goal is to construct the one-sided confidence interval, i.e. calculating the upper bound $C(X)$ similar to Eq.~(\ref{one-sided}). Assume that $\delta$ is an arbitrary real number between 0 and $p$, and $u(m,p,\delta)$ is a real-valued function. Then an upper bound on the tail probability of the binomial distribution can be generically described as
\begin{equation*}
P_X\bigl[ \bar{X} \leq p-\delta \bigl] \leq u(m,p,\delta). \label{tail}
\end{equation*}
Thus, by a straightforward calculation, we can show  
\begin{equation}
P_X\bigl[ p \leq \bar{X}+\delta \bigl] \geq 1-u(m,p,\delta). \label{straight}  
\end{equation}
In Eq.~(\ref{straight}), by setting $\delta$ as $u(m,p,\delta)=\epsilon_{PE}$ for all $p$ and given $m$, we can regard Eq.~(\ref{straight}) as the conservative one-sided confidence interval with confidence level $1-\epsilon_{PE}$, thereby $C(X)=\bar{X}+\delta$ in Eq.~(\ref{one-sided}). Moreover, we can calculate $C(X)$ from the function $u$, the sample size $m$, and the realization of $\bar{X}$ as follows. From the fact that $u(m,p,\delta)=\epsilon_{PE}$ for any $p$, we have
\begin{equation}
u(m,C(X),C(X)-\bar{X}) = \epsilon_{PE}. \label{u}
\end{equation}
By regarding the left-hand side of Eq.~(\ref{u}) as a function of $C(X)$, i.e. $u_{m,\bar{X}}(C(X)):=u(m,C(X),C(X)-\bar{X})$, we get
%\vspace{-2mm} 
\begin{eqnarray}
& &u_{m,\bar{X}}(C(X))=\epsilon_{PE} \label{ce}  \\ 
&\Leftrightarrow& C(X)=u^{-1}_{m,\bar{X}}(\epsilon_{PE}). \label{cx}
\end{eqnarray}
Therefore, we can calculate $C(X)$. Note that the inverse function of $u_{m,\bar{X}}$ exists since it is generally monotonically decreasing function on $[\bar{X},1]$. Furthermore, the tighter the function $u_{m,\bar{X}}$ is, the smaller the value of $C(X)$. Therefore, we can construct a smaller confidence interval by using the tighter bound $u_{m,\bar{X}}$.
%\begin{equation}
%P_X\bigl[ p \leq \bar{X}+u^{-1}_{m,p}(\epsilon_{PE}) \bigl] \geq 1-\epsilon_{PE%}. \label{one-sided-confi}
%\end{equation}
%Sice Eq.~(\ref{one-sided-confi}) is satisfied for all the parameter $p$, $[0,\bar{X}+u^{-1}_{m,p}(\epsilon_{PE})]$ can be interpreted as the conservative confidence interval with confidence revel $1-\epsilon_{PE}$ for $p$. Thus,
%\begin{equation}
%C(X)=\bar{X}+u^{-1}_{m,p}(\epsilon_{PE}). \label{cx}
%\end{equation}
%
%chernoff start
 \item \textit{Chernoff bound\textnormal} \cite{chernoff}\;:\;For any $0\leq\delta\leq p$, Chernoff bound is described by
\begin{equation}
P_X\bigl[ \bar{X} \leq p-\delta \bigl] \leq 2^{-m D(p-\delta||p)}. \label{chernoff}
\end{equation}
%where
%\begin{equation*}
%D(x||y):= x\log_2{\frac{x}{y}}+(1-x)\log_2{\frac{1-x}{1-y}}.
%\end{equation*} \\
By considering $u(m,p,\delta)=2^{-m D(p-\delta||p)}$, we can gain
\begin{equation}
u_{m,\bar{X}}(C(X))=2^{-m D(\bar{X}||C(X))}. \label{cu}
%C(X)=\bar{X}+\phi^{-1}_p(-\log{\epsilon_{PE}}).  \label{cr}
\end{equation}
Thus we can calculate $C(X)$ in the same manner as Eq.~(\ref{cx}).\\
\;\;On the other hand, from Eqs.~(\ref{ce}) and (\ref{cu}), we have
\begin{eqnarray}
&&2^{-m D(\bar{X}||C(X))}=\epsilon_{PE} \nonumber \\
&\Leftrightarrow& D(\bar{X}||C(X))=\log_2{(1/\epsilon_{PE})}/m. \label{aa}
\end{eqnarray}
Moreover, the right-hand side of Eq.~(\ref{aa}) is smallar than $\xi'$, that is, 
\begin{equation*}
\log_2{(1/\epsilon_{PE})}/m < \xi'. 
\end{equation*}
Hence we can see that the confidence interval $[0,C(X)]$ by Chernoff bound is tighter than $\Gamma_{\xi'}$ by comparing Eq.~(\ref{aa}) with Eq.~(\ref{relative}).
%moment start
 \item \textit{Factorial moment bound} \textnormal{\cite{moment}}\;:\;
For any $0<\delta\leq p$, the factorial moment bound is described by
%\begin{equation*}
%P_X\bigl[ \bar{X} \leq p-\delta \bigl] \leq \min_{0\leq n <t}\frac{\mu\{\mu-(1-p)\}\cdots\{\mu-n(1-p)\}}{t(t-1)\cdots(t-n)},
%\end{equation*}
%where $t=m(1-p+\delta)$ and $\mu=m(1-p)$, and the minimization of the right-hand side is achieved when $n^*=\lfloor (t-\mu) / (1-p) \rfloor$. 
\begin{equation}
P_X\bigl[ \bar{X} \leq p-\delta \bigl] \leq \frac{\mu\{\mu-(1-p)\}\cdots\{\mu-n^*(1-p)\}}{t(t-1)\cdots(t-n^*)}, \label{moment-func}
\end{equation}
where $t=m(1-p+\delta)$ and $\mu=m(1-p)$, and $n^*=\lfloor (t-\mu) / p \rfloor$. 
Therefore, by considering
\begin{equation*}
u(m,p,\delta)=\frac{\mu\{\mu-(1-p)\}\cdots\{\mu-n^*(1-p)\}}{t(t-1)\cdots(t-n^*)},
\end{equation*}
we can compute $C(X)$ as well as Chernoff bound.\\
\;\;Since the upper bound in Eq.~(\ref{moment-func}) is tighter than the one in Eq.~(\ref{chernoff}) \cite{philippe}, the value of $u^{-1}_{m,\bar{X}}(\epsilon_{PE})$ , which is calculated from the fractional moment bound is smaller than the one from Chernoff bound, thereby the confidence interval by the fractional moment bound is also tighter.%kalr start
 \item \textit{Klar bound \textnormal{\cite{klar}}}\;:\;Let 
\begin{equation*}
f_x := \left(
\begin{array}{c}
m\\
x
\end{array}
\right)
(1-p)^x p^{m-x} \; (0\leq x \leq m).
\end{equation*}
Then for any $0\leq\delta\leq p$, Klar bound is described by
\begin{equation*}
P_X(\bar{X} \leq p-\delta)\leq \frac{(n+1)p}{n+1-(m+1)(1-p)}f_n, 
\end{equation*}
where $n=m(1-p+\delta)$. Thus, we can calculate $C(X)$ by setting
\begin{equation}
u(m,p,\delta)= \frac{(n+1)p}{n+1-(m+1)(1-p)}f_n. \label{klar-confi}
\end{equation}
\textnormal{
\;In Eq.~(\ref{klar-confi}), if $m$ is very large, it is difficult to compute the binomial coefficient $\left(
\begin{array}{c}
m\\
n
\end{array}
\right)$. %To calculate this value, see Lemma.7 of \cite[p.309]{williams}.}
To calculate this value, we can use the following lemma.}
\end{enumerate}
\begin{lemma}
 \textnormal{\cite[Lemma.7, p.309]{williams}\;
Suppose $m$ and $n(\leq m)$ are integers. Then
\begin{equation*}
\left(
\begin{array}{c}
m\\
n
\end{array}
\right)
\leq \frac{1}{\sqrt{2 \pi m \lambda(1-\lambda)}}2^{m h(\lambda)},
\end{equation*}
where $\lambda=n/m$, and $h(\cdot)$ is the binary entropy.
}
\end{lemma}

%******************************************************************************
%\vspace{-2mm}
\subsection{Computing with the accurate channel estimation}
\label{section-computing}
The computation method of Eve's worst-case ambiguity $\min_{\rho_{AB}\in\Gamma_{\xi}}S_{\rho_{AB}}(X|E)$ in Eq.\ (\ref{lower-bound}) with accurate channel estimation using finite sample bits has not been clarified. Therefore, we show how to compute it in this section.

First of all, observe that the formula (\ref{lower-bound}) found by
Scarani and Renner \cite{PRL,scarani:08} is so general that
we can also just apply Eq.\ (\ref{lower-bound}) to the 
the accurate channel estimation.
There is no need to develop a new analysis for the accurate channel estimation
with finite samples.
So we need to numerically compute $\min_{\rho_{AB}\in\Gamma_{\xi}}S_{\rho_{AB}}(X|E)$
in Eq.\ (\ref{lower-bound}).
However, we use $\Gamma_{\xi'}$ of Eq.\ (\ref{relative}) instead of $\Gamma_{\xi}$. There are two reasons for this choice.
Firstly, $\Gamma_{\xi'}$ is smaller than $\Gamma_{\xi}$ as shown in Section
\ref{subsection-relative-entropy}, and we have
$\min_{\rho_{AB}\in\Gamma_{\xi'}}S_{\rho_{AB}}(X|E) \geq \min_{\rho_{AB}\in\Gamma_{\xi}}S_{\rho_{AB}}(X|E)$.
Secondly, we can differentiate the mathematical expressions in
$\Gamma_{\xi'}$ and the differentiability often helps the numerical
optimization.

An analytical computation of Eve's worst-case ambiguity may be impossible. Therefore, to obtain this value, it is necessary to solve the following minimization problem:
\begin{eqnarray}
\textnormal{minimize} &:& S_{\rho_{AB}}(X|E)\label{opti-prob}\\ 
\textnormal{subject to}&:& \rho_{AB} \textnormal{\;is a real Choi matrix} \nonumber\\
&:& \rho_{AB} \in \Gamma_{\xi'}. \nonumber
\end{eqnarray}
Note that when $\rho_{AB}$ is the real matrix, the optimum value of Eq.~(\ref{opti-prob}) among all the complex Choi matrices is achieved by Proposition 1 of \cite{watanabe:08}.
This allows us to restrict the range of minimization to real matrices.
Without Proposition 1 of \cite{watanabe:08}
the range of minimization must be complex matrices.

%Note also that not $\Gamma_{\xi}$ but $\Gamma_{\xi'}$ is adopted as confidence region in Eq.~(\ref{opti-prob}) because we can calculate $S_{\rho_{AB}}(X|E)$ with $\Gamma_{\xi'}$, and the minimum value of $S_{\rho_{AB}}(X|E)$ within $\Gamma_{\xi'}$ is surely higher than that within $\Gamma_{\xi}$.
Fortunately, this problem is a convex optimization because the objective function $S_{\rho_{AB}}(X|E)$ is a convex with respect to $\rho_{AB}$ \cite{watanabe:08} and $\Gamma_{\xi'}$ is a convex set. Note that the convexity of $\Gamma_{\xi'}$ can be easily proved by facts that a sublevel set of a convex function is convex \cite{boyd} and the relative entropy is convex \cite{cover:01}. 
Hence, we can compute the global optimum value of Eq.~(\ref{opti-prob}).
%******************************************************************************
%\section{Convexity of $\Gamma_{\xi'}$}
%\begin{remark}
%\label{convexity}
%\textnormal{
%In the following, we prove that $\Gamma_{\xi'}$ is a convex set.}
%\begin{proof}
%$\Gamma_{\xi'}$ can be interpreted as the \textit{sublevel set} of the relative entropy. From the fact that a sublevel set of a convex function is convex \cite{boyd} and the relative entropy is convex \cite{cover:01}, $\Gamma_{\xi'}$ is the convex set.
%\end{proof}  
%\end{remark}
\begin{remark}
\textnormal{A standard algorithm to solve a constrained minimization problem like Eq.~(\ref{opti-prob}) is the \textit{interior-point method} (e.g. see \cite{boyd}), and the gradient and the Hessian of the objective function are usually required to use this algorithm. In Eq.~(\ref{opti-prob}), however, it is difficult to calculate those of the objective function $S_{\rho_{AB}}(X|E)$ because $S_{\rho_{AB}}(X|E)$ is the function that depends on eigenvalues of $4\times 4$ matrices. To calculate those derivatives, we can use the method for \textit{spectral functions}. The gradient can be handily derived by using Theorem 1.1 of \cite{lewis}, and the Hessian by Proposition 6.6 of \cite{sendov}.}
\end{remark}

\begin{remark}
\textnormal{The interior-point method requires a \textit{strictly feasible} starting point, which means that the point strictly satisfies all the constraints. In particular, we should find the Choi operator $\rho_{AB}$ satisfied $D(\lambda_m||\lambda_{\infty}(\rho_{AB})) < \xi'$ for given $\lambda_m$ and $\xi'$. Since such a point is not known, we should solve another convex optimization problem,
\begin{eqnarray*}
\textnormal{minimize} &:& D(\lambda_m||\lambda_{\infty}(\rho_{AB})) \\
\textnormal{subject to}&:& \rho_{AB} \textnormal{\;is a real Choi matrix}
\end{eqnarray*}
as a preliminary stage, called \textit{phase I} \cite{boyd}. Note that the starting point of this optimization can be an arbitary Choi matrix. The strictly feasible point found during phase I is then used as the starting point for the original problem, which is called the \textit{phase II}.
}
\end{remark}
%\vspace{-2mm}

\begin{remark}
\textnormal{By switching the role of Alice and Bob in the information reconciliation step, we can sometimes asymptotically gain a higher key rate than the original procedure that is called the direct reconciliation \cite{watanabe:08}. Such a procedure is usually called the reverse reconciliation \cite{boileau05,maurer}.
A non-asymptotical key rate for the reverse reconciliation can be derived by replacing $S_{\rho_{AB}}(X|E)$ of Eq.~(\ref{lower-bound}) with $S_{\rho_{AB}}(Y|E)$ \cite{watanabe:08}. For calculating the gradient and the Hessian of $S_{\rho_{AB}}(Y|E)$, we can use the result in \cite{jankovic}.
}
\end{remark}

\begin{remark}
\normalfont
The optimization problem (\ref{opti-prob}) can also be regarded as
a semidefinite optimization with a nonlinear convex objective function.
Recently, several methods have been proposed for solving such kind
of the optimization problem, for example, \cite{kocvara03,stinglphd,yamashita07,yamashita09}.
In the numerical computation in Section \ref{sec:numcompt},
we used the method proposed in \cite{kocvara03,stinglphd}.
\end{remark}

%******************************************************************************

\subsection{Comparison of Eve's worst-case ambiguities}\label{sec:numcompt}
%\vspace{-1mm}
\label{section-comparison}
The influence from the different channel estimation method appears only in Eve's worst-case ambiguity in Eq.~(\ref{lower-bound}). Therefore, we can compare the secure key rates only by Eve's worst-case ambiguities.

\begin{figure}[t!]
\begin{center}
\includegraphics[width=\linewidth]{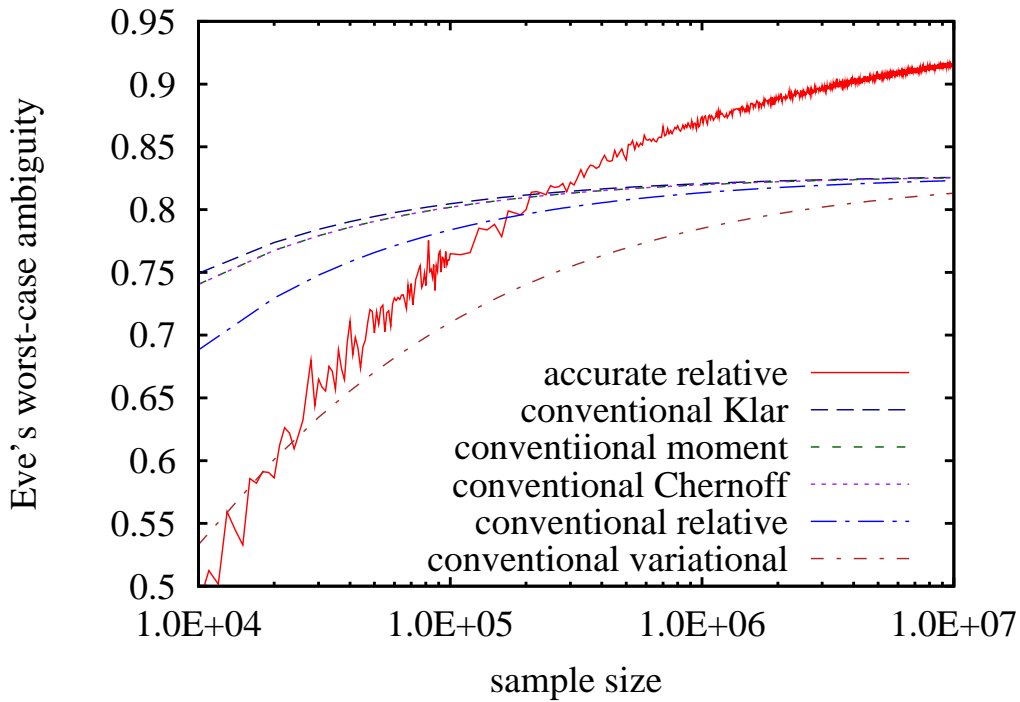}
\end{center}
%\vspace{-2.5mm} 
\caption{{\bf (Color Online)} Comparison of Eve's worst-case ambiguities in the BB84 protocol over the amplitude damping channel against the sample size with the accurate channel estimation. ``accurate relative'' is Eve's worst-case ambiguity with the accurate channel estimation obtained solving the convex optimization Eq.~(\ref{opti-prob}) (see Section \ref{section-computing}). Moreover, ``conventional variational'' and ``conventional relative'' are Eve's worst-case ambiguities with the conventional channel estimation by $\Gamma_{\xi}$ and $\Gamma_{\xi'}$, ``conventional Chernoff,'' ``conventional moment,'' ``conventional Klar'' are ones by the one-sided confidence interval using respective bounds (see Section \ref{section-bino}). Note that ``conventional Chernoff'' and ``conventional Moment'' almost overlap. Parameters are the channel parameter $q=0.1$ (see Eq.~(\ref{amplitude})), $\epsilon_{PE}=10^{-5}$.}\label{fig1}
\end{figure}

\begin{figure}[t!]
\begin{center}
\includegraphics[width=\linewidth]{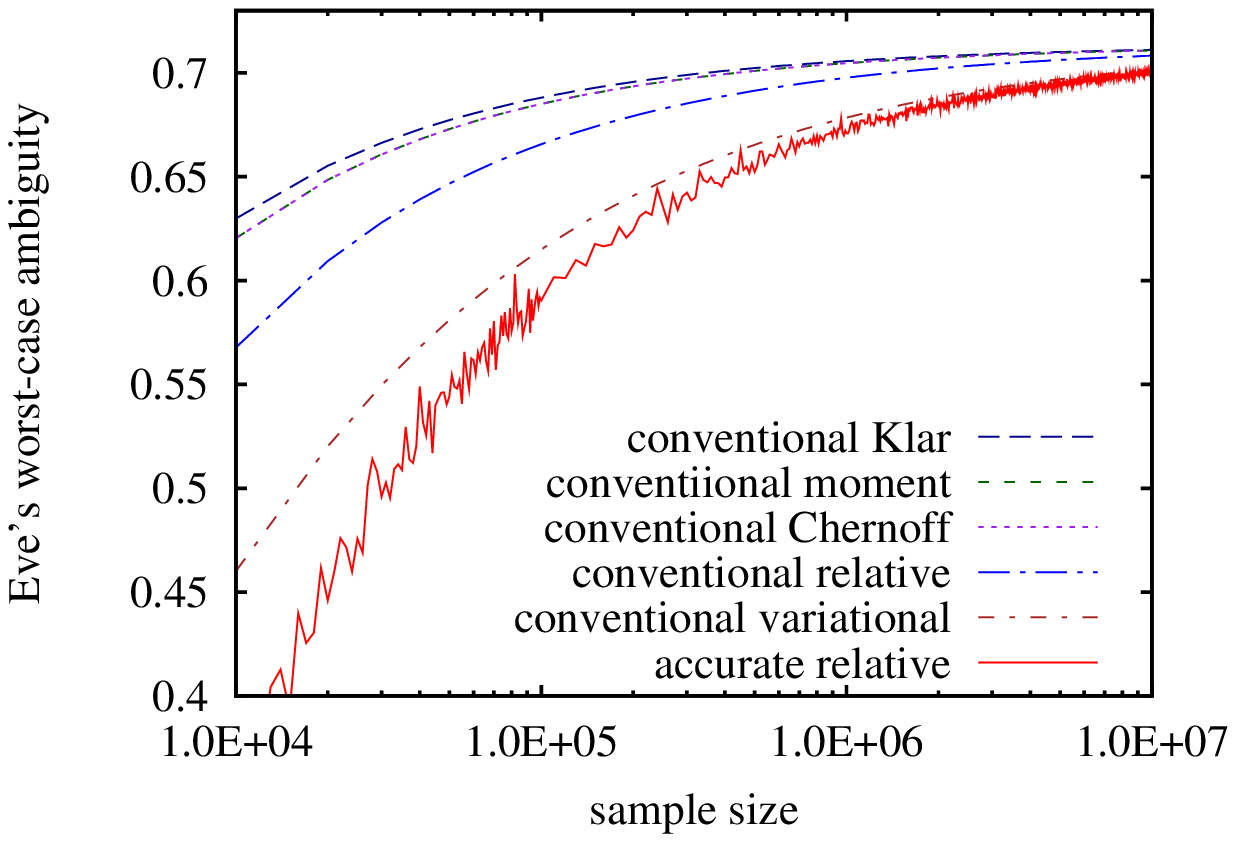}
\end{center}
%\vspace{-2.5mm} 
\caption{{\bf (Color Online)} Comparison over the depolarizing channel. ``accurate relative'' is Eve's worst-case ambiguity with the accurate channel estimation obtained solving the convex optimization Eq.~(\ref{opti-prob}) (see Section \ref{section-computing}). Moreover, ``conventional variational'' and ``conventional relative'' are Eve's worst-case ambiguities with the conventional channel estimation by $\Gamma_{\xi}$ and $\Gamma_{\xi'}$, ``conventional Chernoff'',``conventional moment'',``conventional Klar'' are ones by the one-sided confidence interval using respective bounds (see Section \ref{section-bino}). Note that ``conventional Chernoff'' and ``conventional moment'' almost overlap. Parameters are channel parameter $q=0.1$ (see Eq.~(\ref{depolarizing})), $\epsilon_{PE}=10^{-5}$.}\label{fig2}
%\vspace{-6.5mm}
\end{figure}

In Section \ref{re-confi}, we showed in theory that the confidence interval is smaller in the following order: $\Gamma_{\xi}$, $\Gamma_{\xi'}$, the one-sided confidence interval by using Chernoff bound, the one-sided confidence interval by the factorial moment bound. Therefore, Eve's worst-case ambiguities grow also in this order in the conventional channel estimation. However, the relation between those confidence intervals and the one-sided confidence interval by using Klar bound is not clear. Thus, we compare Eve's worst-case ambiguities in the BB84 protocol by the proposed methods over the following channels:
\begin{enumerate}
%\vspace{-1mm} 
 \item amplitude damping channel
 \begin{eqnarray}
 \label{amplitude}
 \left(
\begin{array}{c}
\theta_Z\\
\theta_X\\
\theta_Y
\end{array}
\right)
\mapsto
\left(\begin{array}{ccc}
1-q&0&0\\
0&\sqrt{1-q}&0\\
0&0&\sqrt{1-q}%
\end{array}\right)
\left(
\begin{array}{c}
\theta_Z\\
\theta_X\\
\theta_Y
\end{array}
\right)+
\left(
\begin{array}{c}
q \\
0\\
0
\end{array}
\right),
\end{eqnarray}
 \item depolarizing channel
  \begin{eqnarray}
  \label{depolarizing}
 \left(
\begin{array}{c}
\theta_Z\\
\theta_X\\
\theta_Y
\end{array}
\right)
\mapsto
\left(\begin{array}{ccc}
1-q&0&0\\
0&1-q&0\\
0&0&1-q%
\end{array}\right)
\left(
\begin{array}{c}
\theta_Z\\
\theta_X\\
\theta_Y
\end{array}
\right),
\end{eqnarray}
\end{enumerate}
where 
$(\theta_Z,\theta_X, \theta_Y)$ describes the representation of a qubit vector in the Bloch sphere, and 
the channel parameter $q$ is a real number between 0 and 1 \cite{nielsen-chuang:00}. Furthermore, we show computation results of Eve's worst-case ambiguity with the accurate channel estimation over those channels on Figs.~\ref{fig1} and \ref{fig2}. These values with the accurate channel estimation are computed by using MATLAB 2009bSP1 and PENNON 1.0, which can be purchased from PENOPT GbR
(\href{http://www.penopt.com}{www.penopt.com}).
We included the MATLAB routines of our numerical computation
into
the supplementary data to this article
so that the scientific community can
verify out results.
Note that the horizontal axis in two figures indicates the sample size used to estimate each channels with the accurate channel method, and the vertical axis indicates Eve's worst-case ambiguity.
%The worst-case Eve's ambiguity are computed with simple size of
%$10,000$, $20,000$, \ldots, $10,000,000$.

%\vspace{-2mm}
\begin{remark}
\textnormal{In the conventional channel estimation, Eve's worst-case ambiguity is calculated as follows. Let $\tilde{p}$ be the worst-case estimate of phase error rate with confidence level $1-\epsilon_{PE}$, namely $C(X)$ in Eq.~(\ref{one-sided}), then Eve's worst-case ambiguity is well-known value $1-h(\tilde{p})$ \cite{renner:05}, where $h(\cdot)$ is a binary entropy.} 
\end{remark}

%\vspace{-3.5mm}
\begin{remark}
\textnormal{The sample size for the accurate channel estimation is about four times as many as that for the conventional channel estimation in our comparison. This is because we estimate the channel by using measurement outcomes only when both Alice and Bob choose $\san{x}$-basis in the conventional channel estimation. While, in contrast, in the accurate channel estimation we estimate the channel by using all measurement outcomes when Alice and Bob choose the transmission basis and the measurement observable among $\san{x}$-basis and $\san{z}$-basis with probability $1/2$ respectively.} 
\end{remark}

%\vspace{-3.5mm}
\begin{remark}
\textnormal{In Figs.~\ref{fig1} and \ref{fig2}, Eve's worst-case ambiguities with the conventional channel estimation are computed by assuming that empirical distribution $\lambda_m$ is equal to theoretical distribution determined uniquely by $\rho_{AB}$ of each channel, because the channel corresponding to any $\lambda_m$ always exists. By contrast, in the accurate channel estimation, the channel corresponding to measured statistic $\lambda_m$ does not necessarily exist \cite{ziman:05}. Thus we compute Eve's worst-case ambiguities by $\lambda_m$ generated with the pseudo-random number generator, to keep fairness of the comparisons.}
\end{remark}

%\vspace{-2mm}

%\newpage

\subsection{Discussion}
From Figs.~\ref{fig1} and \ref{fig2}, we can see two facts: First, our proposed confidence intervals improve non-asymptotically Eve's worst-case ambiguity over the conventional confidence region. The amount of the improvement by ``conventional Klar'' compared with ``conventional variational'' is about $1.1$\% at $10^7$ samples in both figures. Klar bound is the larger than Chernoff bound and the factorial moment bound, though  the differences are small. In addition, since convergences of these bounds are faster than that by $\Gamma_{\xi}$, we can gain a higher key rate for fewer samples in which key rate with $\Gamma_{\xi}$ is small. For example from Fig. ~2, when the sample size is $10^4$, we can see that the value by $\Gamma_{\xi}$ is about $0.56$, in contrast, the value by Klar bound is about $0.67$. Secondly, Eve's worst-case ambiguity with the accurate channel estimation is non-asymptotically much higher compared with all values with the conventional channel estimation over the amplitude damping channel, for example from Fig. ~1, when the sample size is $10^7$, about $20\%$ higher than $\Gamma_{\xi}$. However, from Fig. ~2, the accurate estimate is the smallest over the depolarizing channel.

Observe that the accurate channel estimation (with relative entropy)
gives the worse estimate than
the conventional channel estimation with relative entropy,
for all samples sizes over the depolarizing channels,
though their asymptotic limits of $\min S_{\rho_{AB}}(X|E)$
are the same as shown in \cite{watanabe:08}.
In the authors' opinion, this is because 
the accurate channel estimation has to estimate the larger number 
of parameters and the accuracy of estimate is degraded by the increase
in the number of parameters.
Note that the number of parameters is 7 for the accurate channel
estimation and 1 for the conventional one.
On the other hand, the accurate channel estimation gives better estimates
with larger sample sizes over the amplitude damping channel.
This is because the asymptotic limit of the accurate channel
estimation is much larger than the conventional one,
as shown in \cite{watanabe:08},
while the accurate channel estimation also experiences the degradation by
the increased number of parameters with smaller sample sizes as well.
Since the asymptotic limit of the accurate channel estimation is
always larger than the conventional one if the channel is not
Pauli one \cite{watanabe:09},
the accurate channel estimation is expected to work better when
the channel is supposed to be non-Pauli and the sample size for
channel estimation is large.

%******************************************************************************
%******************************************************************************
%\vspace{-2mm}
\section{Conclusion}
%\vspace{-2mm}
\label{section-conclusion}
The accurate channel method non-asymptotically increases the key rate over the amplitude damping channel. Thus, we should not discard mismatched measurement outcomes in that case. However, the key rate non-asymptotically depreciates over the depolarizing channel. On the other hand, in the conventional channel estimation, the non-asymptotic key rate shown by Scarani \textit{et al.}\ is improved by reconstructing the confidence interval for a channel using the one-sided interval estimation with tail probability bounds. One-sided intervals can improve the key rate in the following order of tail probability bounds: the variational distance, the relative entropy, Chernoff bound, factorial moment bound, Klar bound.
%******************************************************************************
%******************************************************************************
%\vspace{-2mm}
\section*{Acknowledgment}

The authors greatly appreciate critical comments by the referee that 
improved the presentation of this paper very much.
The first author would like to give heartful thanks to Dr.~Shun Watanabe whose enormous support and insightful comments were invaluable. The second author would like to thank Dr.~Anthony Leverrier, Prof.~Valerio Scarani, and Prof.~Renato Renner for helpful discussions, and Prof.\ Michael Stingl for helping us to use the PENNON optimizer.
This research was partly supported
by the Japan Society for the Promotion of Science
under Grants-in-Aid for Young Scientists No.\ 22760267.
%******************************************************************************
%******************************************************************************

%\bibliographystyle{habbrv-href}
%\bibliography{mrabbrev,sano}

\begin{thebibliography}{10}

\bibitem{ben:04}
M.~Ben-Or, M.~Horodecki, D.~W. Leung, D.~Mayers, and J.~Oppenheim.
\newblock The universal composable security of quantum key distribution.
\newblock In J.~Kilian, editor, {\em Proc. Second Theory of Cryptography
  Conference, TCC 2005}, volume 3378 of {\em Lecture Notes in Computer
  Science}, pages 386--406. Springer-Verlag, Feb. 2005.
\newblock \href{http://dx.doi.org/10.1007/b106171}{\path{doi:10.1007/b106171}}.

\bibitem{bb84}
C.~H. Bennett and G.~Brassard.
\newblock Quantum cryptography: Public key distribution and coin tossing.
\newblock In {\em Proc.\ IEEE Intl.\ Conf.\ on Computers, Systems, and Signal
  Processing}, pages 175--179, 1984.

\bibitem{boileau05}
J.-C. Boileau, J.~Batuwantudawe, and R.~Laflamme.
\newblock Higher-security thresholds for quantum key distribution by improved
  analysis of dark counts.
\newblock {\em Phys.\ Rev.\ A}, 72(3):032321, Sept. 2005.
\newblock arXiv:quant-ph/0502140,
  \href{http://dx.doi.org/10.1103/PhysRevA.72.032321}{\path{doi:10.1103/PhysRe%
vA.72.032321}}.

\bibitem{boyd}
S.~Boyd and L.~Vandenberghe.
\newblock {\em Convex Optimization}.
\newblock Cambridge University Press, 2004.

\bibitem{cai:08}
R.~Y.~Q. Cai and V.~Scarani.
\newblock Finite-key analysis for practical implementations of quantum key
  distribution.
\newblock {\em New. J. Phys.}, 11(4):045024, Apr. 2009.
\newblock arXiv:0811.2628,
  \href{http://dx.doi.org/10.1088/1367-2630/11/4/045024}{\path{doi:10.1088/136%
7-2630/11/4/045024}}.

\bibitem{casella-berger:01}
G.~Casella and R.~L. Berger.
\newblock {\em Statistical Inference}.
\newblock Duxbury Press, 2nd edition, 2001.

\bibitem{chernoff}
H.~Chernoff.
\newblock A measure of asymptotic efficiency for tests of a hypothesis based on
  the sum of observations.
\newblock {\em Ann. Math. Statist.}, 23(4):493--507, 1952.
\newblock
  \href{http://dx.doi.org/10.1214/aoms/1177729330}{\path{doi:10.1214/aoms/1177%
729330}}.

\bibitem{choi}
M.-D. Choi.
\newblock Completely positive linear maps on complex matrices.
\newblock {\em Linear Algebra and Appl.}, 10(3):285--290, June 1975.
\newblock
  \href{http://dx.doi.org/10.1016/0024-3795(75)90075-0}{\path{doi:10.1016/0024%
-3795(75)90075-0}}.

\bibitem{cover:01}
T.~M. Cover and J.~A. Thomas.
\newblock {\em Elements of Information Theory}.
\newblock Wiley Interscience, 2nd edition, 2006.

\bibitem{donna}
D.~Dodson et~al.
\newblock Updating quantum cryptography report ver. 1, May 2009.
\newblock arXiv:0905.4325.

\bibitem{jankovic}
M.~S. Jankovic.
\newblock Exact $n$th derivatives of eigenvalues and eigenvectors.
\newblock {\em Journal of Guidance, Control, and Dynamics}, 17(1):136--144,
  Jan. 1994.
\newblock \href{http://dx.doi.org/10.2514/3.21170}{\path{doi:10.2514/3.21170}}.

\bibitem{klar}
B.~Klar.
\newblock Bounds on tail probabilities of discrete distributions.
\newblock {\em Probability in the Engineering and Informational Sciences},
  14(2):161--171, Apr. 2000.
\newblock
  \href{http://dx.doi.org/10.1017/S0269964800142032}{\path{doi:10.1017/S026996%
4800142032}}.

\bibitem{kocvara03}
M.~Ko\v{c}vara and M.~Stingl.
\newblock {PENNON}: a code for convex nonlinear and semidefinite programming.
\newblock {\em Optimization Methods and Software}, 18(3):317--333, June 2003.
\newblock
  \href{http://dx.doi.org/10.1080/1055678031000098773}{\path{doi:10.1080/10556%
78031000098773}}.

\bibitem{leverrier:09}
A.~Leverrier.
\newblock {\em Theoretical study of continuous-variable quantum key
  distribution}.
\newblock PhD thesis, Telecom ParisTech, Paris, France, 2009.
\newblock Available from: \url{http://www.infres.enst.fr/~leverrie/}.

\bibitem{leverrier10}
A.~Leverrier, F.~Grosshans, and P.~Grangier.
\newblock Finite-size analysis of a continuous-variable quantum key
  distribution.
\newblock {\em Phys.\ Rev.\ A}, 81(6):062343, June 2010.
\newblock arXiv:1005.0339,
  \href{http://dx.doi.org/10.1103/PhysRevA.81.062343}{\path{doi:10.1103/PhysRe%
vA.81.062343}}.

\bibitem{lewis}
A.~S. Lewis.
\newblock Derivatives of spectral functions.
\newblock {\em Mathematics of Operations Research}, 21(3):576--588, Aug. 1996.
\newblock
  \href{http://dx.doi.org/10.1287/moor.21.3.576}{\path{doi:10.1287/moor.21.3.5%
76}}.

\bibitem{williams}
F.~J. MacWilliams and N.~J.~A. Sloane.
\newblock {\em The Theory of Error-Correcting Codes}.
\newblock Elsevier, Amsterdam, 1977.

\bibitem{maurer}
U.~Maurer.
\newblock Secret key agreement by public discussion from common information.
\newblock {\em IEEE Trans.\ Inform.\ Theory}, 39(3):733--742, May 1993.
\newblock
  \href{http://dx.doi.org/10.1109/18.256484}{\path{doi:10.1109/18.256484}}.

\bibitem{philippe}
P.~Naveau.
\newblock Comparison between the {Chernoff} and factorial moment bounds for
  discrete random variables.
\newblock {\em The American Statistician}, 51(1):40--41, Feb. 1997.
\newblock \href{http://dx.doi.org/10.2307/2684691}{\path{doi:10.2307/2684691}}.

\bibitem{nielsen-chuang:00}
M.~A. Nielsen and I.~L. Chuang.
\newblock {\em Quantum Computation and Quantum Information}.
\newblock Cambridge University Press, Cambridge, UK, 2000.

\bibitem{moment}
T.~K. Philips and R.~Nelson.
\newblock The moment bound is tighter than {Chernoff's} bound for positive tail
  probabilities.
\newblock {\em The American Statistician}, 49(2):175--178, May 1995.
\newblock \href{http://dx.doi.org/10.2307/2684633}{\path{doi:10.2307/2684633}}.

\bibitem{renner:05}
R.~Renner, N.~Gisin, and B.~Kraus.
\newblock Information-theoretic security proof for quantum-key-distribution
  protocols.
\newblock {\em Phys.\ Rev.\ A}, 72(1):012332, July 2005.
\newblock arXiv:quant-ph/0502064,
  \href{http://dx.doi.org/10.1103/PhysRevA.72.012332}{\path{doi:10.1103/PhysRe%
vA.72.012332}}.

\bibitem{renner:04}
R.~Renner and R.~K\"onig.
\newblock Universally composable privacy amplification against quantum
  adversaries.
\newblock In J.~Kilian, editor, {\em Proc. Second Theory of Cryptography
  Conference, TCC 2005}, volume 3378 of {\em Lecture Notes in Computer
  Science}, pages 407--425. Springer-Verlag, Feb. 2005.
\newblock \href{http://dx.doi.org/10.1007/b106171}{\path{doi:10.1007/b106171}}.

\bibitem{PRL}
V.~Scarani and R.~Renner.
\newblock Quantum cryptography with finite resources: Unconditional security
  bound for discrete-variable protocols with one-way postprocessing.
\newblock {\em Phys.\ Rev.\ Lett.}, 100(20):200501, May 2008.
\newblock arXiv:0708.0709,
  \href{http://dx.doi.org/10.1103/PhysRevLett.100.200501}{\path{doi:10.1103/Ph%
ysRevLett.100.200501}}.

\bibitem{scarani:08}
V.~Scarani and R.~Renner.
\newblock Security bounds for quantum cryptography with finite resources.
\newblock In Y.~Kawano and M.~Mosca, editors, {\em Theory of Quantum
  Computation, Communication, and Cryptography}, volume 5106 of {\em Lecture
  Notes in Computer Science}, pages 83--95. Springer-Verlag, Nov. 2008.
\newblock arXiv:0806.0120,
  \href{http://dx.doi.org/10.1007/978-3-540-89304-2_8}{\path{doi:10.1007/978-3%
-540-89304-2_8}}.

\bibitem{sendov}
H.~S. Sendov.
\newblock The higher-order derivatives of spectral functions.
\newblock {\em Linear Algebra and Appl.}, 424(1):240--281, July 2007.
\newblock
  \href{http://dx.doi.org/10.1016/j.laa.2006.12.013}{\path{doi:10.1016/j.laa.2%
006.12.013}}.

\bibitem{shor}
P.~W. Shor and J.~Preskill.
\newblock Simple proof of security of the {BB84} quantum key distribution
  protocol.
\newblock {\em Phys.\ Rev.\ Lett.}, 85(2):441--444, July 2000.
\newblock arXiv:quant-ph/0003004,
  \href{http://dx.doi.org/10.1103/PhysRevLett.85.441}{\path{doi:10.1103/PhysRe%
vLett.85.441}}.

\bibitem{stinglphd}
M.~Stingl.
\newblock {\em On the Solution of Nonlinear Semidenite Programs by Augmented
  {Lagrangian} Methods}.
\newblock PhD thesis, University of Erlangen-N\"urnberg, 2006.
\newblock Available from: \url{http://www.am.uni-erlangen.de/~kocvara/pennon/}.

\bibitem{watanabe:09}
S.~Watanabe.
\newblock {\em A Study of Channel Estimation and Postprocessing in Quantum Key
  Distribution Protocols}.
\newblock PhD thesis, Tokyo Institute of Technology, Tokyo, Japan, Mar. 2009.
\newblock arXiv:0904.4083.

\bibitem{watanabe:08}
S.~Watanabe, R.~Matsumoto, and T.~Uyematsu.
\newblock Tomography increases key rates of quantum-key-distribution protocols.
\newblock {\em Phys.\ Rev.\ A}, 78(4):042316, Oct. 2008.
\newblock arXiv:0802.2419,
  \href{http://dx.doi.org/10.1103/PhysRevA.78.042316}{\path{doi:10.1103/PhysRe%
vA.78.042316}}.

\bibitem{yamashita09}
H.~Yamashita and H.~Yabe.
\newblock Local and superlinear convergence of a primal-dual interior point
  method for nonlinear semidefinite programming.
\newblock Optimization Online eprint, Aug. 2009.
\newblock Available from:
  \url{http://www.optimization-online.org/DB_HTML/2009/08/2366.html}.

\bibitem{yamashita07}
H.~Yamashita, H.~Yabe, and K.~Harada.
\newblock A primal-dual interior point method for nonlinear semidefinite
  programming.
\newblock Optimization Online eprint, June 2007.
\newblock Available from:
  \url{http://www.optimization-online.org/DB_HTML/2007/06/1692.html}.

\bibitem{ziman:05}
M.~Ziman, M.~Plesch, V.~Bu\v{z}ek, and P.~\v{S}telmachovi\v{c}.
\newblock Process reconstruction: From unphysical to physical maps via maximum
  likelihood.
\newblock {\em Phys.\ Rev.\ A}, 72(2):022106, Aug. 2005.
\newblock
  \href{http://dx.doi.org/10.1103/PhysRevA.72.022106}{\path{doi:10.1103/PhysRe%
vA.72.022106}}.

\end{thebibliography}

\end{document}